\documentclass[sigconf]{acmart}

\usepackage{multirow}
\usepackage{adjustbox}
\usepackage{balance}
\usepackage{makecell}
\usepackage{bm}
\usepackage{enumitem}
\usepackage{algorithm}
\usepackage{algorithmicx,tikz}
\usepackage{algcompatible}

\usepackage{amsmath,amsfonts}
\usepackage{array}
\usepackage{graphicx}
\usepackage{textcomp}

\AtBeginMaketitle{%

\def\EDBTISSN{2367-2005}
\def\EDBTISBN{978-3-98318-104-9}
\setcopyright{none}
\copyrightyear{\copyright\ 2026 Copyright held by the owner/author(s).
  Published on \url{OpenProceedings.org} under ISBN \EDBTISBN, series ISSN \EDBTISSN.
  Distribution of this paper is permitted under the terms of the
  Creative Commons license CC-by-nc-nd 4.0}
\acmYear{2026}
\acmDOI{}
\acmISBN{}

\acmConference[EDBT '26]{Extending Database Technology}{24-27 March 2026}{Tampere (Finland)}

\settopmatter{printacmref=false, printccs=false, printfolios=false}

\newsavebox{\ximagebox}
\newlength{\ximageheight}
\newsavebox{\xglyphbox}
\newlength{\xglyphheight}
\newcommand{\xbox}[1]%
  {\savebox{\ximagebox}{#1}%
  \settoheight{\ximageheight}{\usebox{\ximagebox}}%
  \savebox{\xglyphbox}{\color{white}\char32}%
  \settoheight{\xglyphheight}{\usebox{\xglyphbox}}%
  \raisebox{\ximageheight}[0pt][0pt]{\raisebox{-\xglyphheight}[0pt][0pt]{%
    \makebox[0pt][l]{\usebox{\xglyphbox}}}}%
    \usebox{\ximagebox}%
    \raisebox{0pt}[0pt][0pt]{\makebox[0pt][r]{\usebox{\xglyphbox}}}}

\newsavebox{\LogoBox}
\sbox{\LogoBox}{\includegraphics[height=1cm]{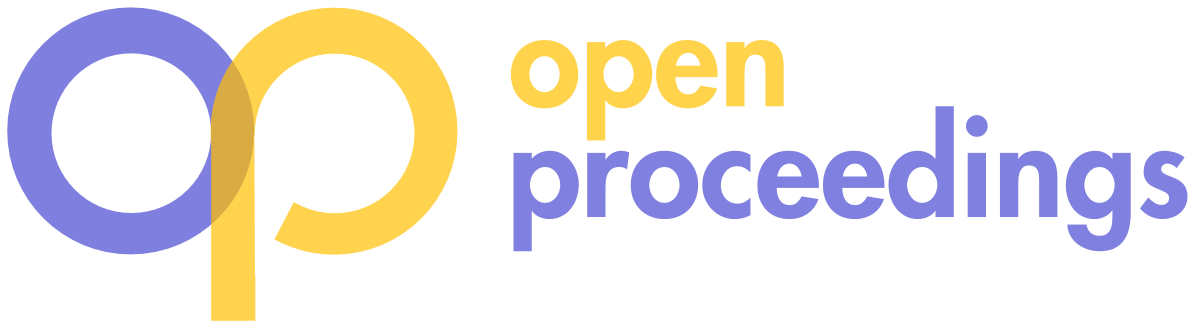}}

\fancypagestyle{firstpagestyle}{%
  \fancyhf{}%
  \fancyfoot{}%
  \fancyhead[L,C]{}
  \fancyhead[R]{\href{https://OpenProceedings.org/}{\raisebox{15pt}[0pt][0pt]{\xbox{\usebox{\LogoBox}}}}}
  }

\hypersetup{%
  pdfcopyright={CC-by-nc-nd},
  pdfpublisher={OpenProceedings.org -- University of Konstanz, Germany},
  pdfcreator={LaTeX with acmart, hyperref, and EDBT modifications},
  pdfvolumenum={29},
  pdfissuenum={2},
  pdfisbn={\EDBTISBN},
  pdfissn={\EDBTISSN}
}

}

\begin{document}


\title{Efficient Retrieval Scaling with Hierarchical Indexing for\\ Large Scale Recommendation}



\author{Dongqi Fu}
\email{dongqifu@meta.com}
\orcid{0000-0002-8726-9234}
\affiliation{%
  \institution{Meta}
  \country{USA}
}
\author{Kaushik Rangadurai}
\email{krangadu@meta.com}
\affiliation{%
  \institution{Meta}
  \country{USA}
}
\author{Haiyu Lu}
\email{hylu@meta.com}
\affiliation{%
  \institution{Meta}
  \country{USA}
}
\author{Yunchen Pu}
\email{pyc40@meta.com}
\affiliation{%
  \institution{Meta}
  \country{USA}
}
\author{Siyang Yuan}
\email{syyuan@meta.com}
\affiliation{%
  \institution{Meta}
  \country{USA}
}
\author{Minhui Huang}
\email{mhhuang@meta.com}
\affiliation{%
  \institution{Meta}
  \country{USA}
}
\author{Yiqun Liu}
\email{yiqliu@meta.com}
\affiliation{%
  \institution{Meta}
  \country{USA}
}
\author{Golnaz Ghasemiesfeh}
\email{golnazghasemi@meta.com}
\affiliation{%
  \institution{Meta}
  \country{USA}
}
\author{Xingfeng He}
\email{xingfenghe@meta.com}
\affiliation{%
  \institution{Meta}
  \country{USA}
}
\author{Fangzhou Xu}
\email{fxu@meta.com}
\affiliation{%
  \institution{Meta}
  \country{USA}
}
\author{Andrew Cui}
\email{andycui97@meta.com}
\affiliation{%
  \institution{Meta}
  \country{USA}
}
\author{Vidhoon Viswanathan}
\email{vidhoon@meta.com}
\affiliation{%
  \institution{Meta}
  \country{USA}
}
\author{Lin Yang}
\email{ylin1@meta.com}
\affiliation{%
  \institution{Meta}
  \country{USA}
}
\author{Liang Wang}
\email{liangwang@meta.com}
\affiliation{%
  \institution{Meta}
  \country{USA}
}
\author{Jiyan Yang}
\email{chocjy@meta.com}
\affiliation{%
  \institution{Meta}
  \country{USA}
}
\author{Chonglin Sun}
\email{clsun@meta.com}
\affiliation{%
  \institution{Meta}
  \country{USA}
}



\renewcommand{\shortauthors}{Dongqi Fu et al.} 

\begin{abstract}
The increase in data volume, computational resources, and model parameters during training has led to the development of numerous large-scale industrial retrieval models for recommendation tasks. However, effectively and efficiently deploying these large-scale foundational retrieval models remains a critical challenge that has not been fully addressed. Common quick-win solutions for deploying these massive models include relying on offline computations (such as cached user dictionaries) or distilling large models into smaller ones. Yet, both approaches fall short of fully leveraging the representational and inference capabilities of foundational models.
In this paper, we explore whether it is possible to learn a hierarchical organization over the memory of foundational retrieval models. Such a hierarchical structure would enable more efficient search by reducing retrieval costs while preserving exactness. To achieve this, we propose jointly learning a hierarchical index using cross-attention and residual quantization for large-scale retrieval models. We also present its real-world deployment at Meta, supporting daily advertisement recommendations for billions of Facebook and Instagram users.
Interestingly, we discovered that the intermediate nodes in the learned index correspond to a small set of high-quality data. Fine-tuning the model on this set further improves inference performance, and concretize the concept of "test-time training" within the recommendation system domain.
We demonstrate these findings using both internal and public datasets with strong baseline comparisons and hope they contribute to the community’s efforts in developing the next generation of foundational retrieval models.
\end{abstract}

\keywords{Foundation Retrieval Model, Hierarchical Index}

\maketitle

\section{Introduction}
With the success of foundation models, substantial research has focused on exploring scaling laws~\cite{DBLP:journals/corr/abs-2001-08361, DBLP:journals/corr/abs-2203-15556}, such as those involving data volume, computational resources, and model parameters during training, with the aim of achieving improved performance at inference time.
Building upon the scaling advancements in foundation language models, recent research has extended to foundation ranking and retrieval models for recommendation ~\cite{DBLP:journals/corr/abs-2208-08489, DBLP:conf/aaai/ShinKKRJ0K23, DBLP:conf/recsys/ZhangHLCZW24, DBLP:conf/sigir/FangZAMS0024, DBLP:journals/corr/abs-2412-00714}, catalyzing the development of numerous large-scale industrial recommendation systems such as Wukong~\cite{DBLP:conf/icml/ZhangLCNLLZHYWP24}, HSTU~\cite{DBLP:conf/icml/ZhaiLLWLCGGGHLS24}, InterFormer~\cite{DBLP:journals/corr/abs-2411-09852}, and ExFM~\cite{DBLP:conf/www/LiangL0LSZZCZL025}.

\begin{figure*}[t]
  \centering
  \includegraphics[width=0.79\textwidth]{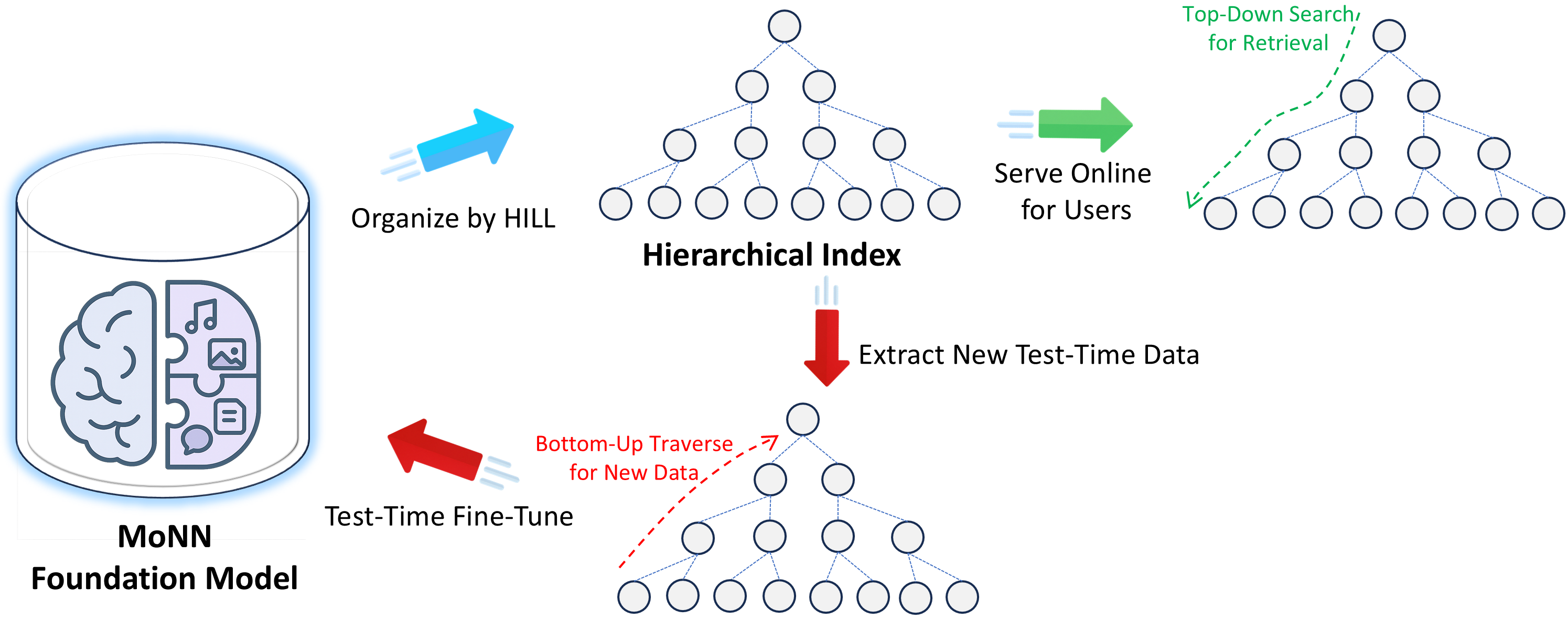}
   \caption{Overall Pipeline of MoNN Foundation Retrieval Model with HILL Index.}
  \label{fig:framework}
\end{figure*}

However, given the surging development of enormous foundation models for ranking and retrieval in recommendation, how to deploy them to serve real-world (e.g., high frequency and heavy load) scenarios well is underexplored or not fully discussed. The most appealing concern can be the inference and retrieval cost.
To leverage the inference ability of foundation retrieval models, passing the user and item features into the large-scale complex retrieval models to get the preference score prediction is often time-consuming and costly, especially when facing a massive volume of user-item pairs.
For the above case, the quick win solution can include: (1) pre-computing the retrieval set for users offline and serving the static result for a certain time window or (2) distilling the large models into small models and shipping them; but neither solutions fully release the representation and inference ability of large models.

In addition, previous related research shows that using an index structure to organize items such that approximately searching over this index (e.g., beam search) can reduce the search space and fast output similar pairs~\cite{DBLP:conf/kdd/ZhuLZLHLG18, gao2020deep, DBLP:conf/nips/FengLLLC22, DBLP:conf/sigir/LiAZM0LC23, liu2024learning, DBLP:conf/iclr/Li0LLBY00G25}, but not fit today's large-scale foundation model scenario in industry, which usually adopts deep and complex-connected neural networks to learn sophisticated interaction between users and items with rich structured descriptions, to the best of our knowledge.
Hence, it motivates our research to learn a hierarchical index over the memory of large-scale foundation retrieval models, such that just searching this exactness-aware hierarchical index can output the retrieval result fast by avoiding much unnecessary search space.

Therefore, in this paper, we first share our experience in deploying large-scale foundation retrieval models at Meta Ads Platform, i.e., learning the hierarchical index for foundation retrieval models helps it obtain an effective-efficiency-balanced stage.
To be specific, we introduce our \textbf{Hierarchical Index Learning} method, called \textbf{HILL}, which aims to hierarchically organize the memory in the foundation retrieval model, such that the searching and retrieval along the structure can be speed up and maintain the exactness. To achieve this goal, HILL leverages the cross-attention mechanism and residual quantization learning and enjoys the co-training with the foundation retrieval model. 
Moreover, to emphasize the reproducibility, we also first time systematically decipher the foundation retrieval model at Meta, called \textbf{Modular Neural Network} (\textbf{MoNN}), which is currently in service for recommending advertisements to daily Facebook and Instagram users, in terms of neural architecture, training procedures, and loss functions.

Interestingly, we also found that the intermediate-level nodes in the built hierarchical index tree compose a small (compared to the volume of users and items) but high-quality data source, which can fine-tune the pre-trained retrieval model to achieve a "\textit{test-time training}" inference upgrade~\cite{DBLP:conf/icml/SunWLMEH20, DBLP:conf/iclr/Hardt024, yuksekgonul2026learning}.
In short, test-time training (TTT) refers to methods that update model parameters during inference and typically does not require ground-truth labels, which is primarily designed to improve and adapt the model rather than relying solely on pretraining.

The entire pipeline of MoNN and HILL learning the index for online service and extracting the new data pair for test-time training is shown in Figure~\ref{fig:framework}, and the rest of the paper is organized as follows: In Section~\ref{sec: foundation model}, we introduce the MoNN foundation model deployed at Meta for user advertisement retrieval to pave the way for introducing the hierarchical index learning method in Section~\ref{sec: index learning}. Starting from Section~\ref{sec: experiment}, we show various experiments, including offline public dataset benchmark performance and online deployment service. After we discuss related work in Section~\ref{sec: related work}, we finally conclude the paper with several future directions in Section~\ref{sec: conclusion}.



\section{Foundation Retrieval Model at Meta - MoNN}
\label{sec: foundation model}
In this section, we briefly introduce the architecture of the foundation model deployed in Meta for retrieval, which is called Modular Neural Network (MoNN), and is flexible to operate under different infrastructure constraints.

\begin{figure}[h]
  \centering
  \includegraphics[width=0.4\textwidth]{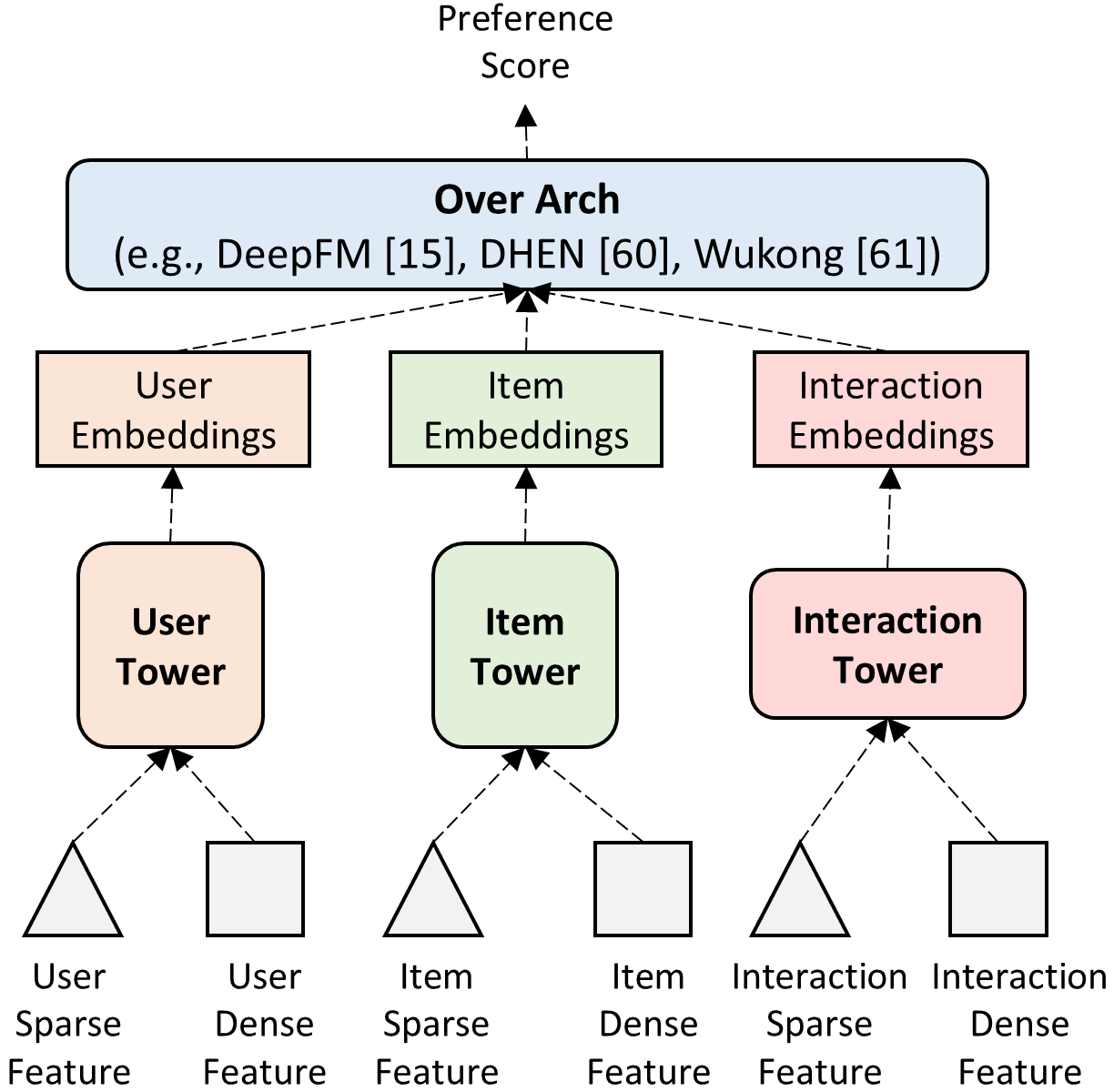}
  \caption{A MoNN Block}
  \label{fig:MoNN}
\end{figure}

In general, Modular Neural Network (MoNN) enhances the learning of sophisticated user and item interactions beyond a single dot product while maintaining high efficiency. As shown in Figure~\ref{fig:MoNN}, MoNN has a modularized design comprising separate modules for user representations (\textit{User Tower}), item representations (\textit{Item Tower}), and the interaction between user and item (\textit{Interaction Tower}).

\textbf{User Tower}.
In short, the user tower takes user features to generate fixed-size user embeddings.
These features can be dense (e.g., number of clicks by the user) and sparse (e.g., user engaged videos).
Sparse features are processed by an embedding lookup table, and then all feature embeddings are concatenated and fed into the tower.
Given that the user tower only needs to be computed once and shared across a vast number of items, it could scale up to very high complexity.

\textbf{Item Tower}.
Similarly, the item tower mirrors the user tower, processing item dense (e.g., item historical click-through rate) and sparse features (e.g., content of item). Sparse features are input into another embedding lookup table, and all feature outputs are concatenated and fed into the tower.

\textbf{Interaction Tower}.
The interaction tower operates on <user, item> interaction features (dense and sparse) as input. It follows a similar architecture to the user tower and item tower to produce the corresponding embeddings.
To be more specific, the interaction tower is computationally intensive as it runs for each pair of user and item. To minimize the computation cost for <user, item> interaction features, we propose to use the \textit{Inverted Index Based Interaction Features} (I2IF), where an inverted index is employed for indexing item information with user information written as a query to perform efficient crossing computation. 

\textbf{Over Architecture}.
Sitting atop the three underlying towers is called over architecture, short for \textbf{OverArch}, which is responsible for aggregating all information comprehensively and producing the user item preference score. OverArch can leverage DHEN~\cite{zhang2022dhendeephierarchicalensemble} (or DeepFM~\cite{DBLP:conf/ijcai/GuoTYLH17}, or Wukong~\cite{DBLP:journals/corr/abs-2203-11014}) to generate numerical logits.

\textbf{Training Setup}.
Briefly, MoNN model is trained on a large-scale training dataset with clicks and conversions as labels, impressions (non-click or conversion) as negatives, and additional unlabeled data used for semi-supervised learning to debias the model.
A wide range of features (e.g., $O(1000)$) is used as input, and MoNN is optimized for multiple tasks, e.g., the click task and the conversion task.
Hence, MoNN is then trained using a multi-task cross-entropy loss $\mathcal{L}$ as follows
\begin{equation}
    \mathcal{L} = \mathcal{L}_{sup} + \mathcal{L}_{unsup}
\end{equation}
and the supervised loss function $\mathcal{L}_{sup}$ is expressed as 
\begin{equation}
\mathcal{L}_{sup} = \frac{-1}{S}\sum_{i = 1}^S\sum_{t = 1}^T w_t(y_{ti}log(\hat{y}_{ti}) + (1-y_{ti}) (log(1-\hat{y}_{ti})))
\end{equation}
where $w_t$ is the weight for task $t$, $t \in \{1,2, \dots, T\}$, representing its importance.
$y_{ti} \in \{0, 1\}$ is the ground-truth label for sample $i$ in task $t$, $\hat{y}_{ti}$ is the predicted value of the model for sample $i$ in task $t$, and $S$ is the number of samples. Similarly, the unsupervised loss function $\mathcal{L}$ is expressed as
\begin{equation}
\mathcal{L}_{unsup} = \frac{-1}{S}\sum_{i = 1}^S\sum_{t = 1}^T distill(\hat{y}_{ti}, y^{model}_{ti})
\end{equation}
where $y^{model}_{ti}$ is the soft label generated by MoNN or another well-trained teacher model, and $distill$ function can be instanced also as the cross-entropy.

\section{Hierarchical Index Learning (HILL)}
\label{sec: index learning}
To avoid the tree construction process relying solely on item embeddings to lose information~\cite{DBLP:conf/kdd/ZhuLZLHLG18, gao2020deep, DBLP:conf/nips/FengLLLC22, DBLP:conf/sigir/LiAZM0LC23, liu2024learning, DBLP:conf/iclr/Li0LLBY00G25}, but retaining as much information as possible, we aim to design a learning-based tree construction method. Moreover, given the comprehensiveness of MoNN to take ample <user, item> information, we then design a joint learning method for the hierarchical index construction.

Formally, in this section, we introduce our Hierarchical Index Learning method, named HILL, which can build up the hierarchical structure to organize items to help the foundation retrieval model (like MoNN) to retrieve the most relevant items for users, and also produce a small amount of high-quality new data to fine-tune the model for test-time training effectiveness.

First, we briefly introduce the overview of HILL in Section~\ref{sec: overview}. Stepping into details, we then introduce how to learn one layer mapping function in Section~\ref{sec: one-layer}, how to residually stack up layers in Section~\ref{sec: cross-layer}, how to optimize the learning process in Section~\ref{sec: optimization}, an approximation learning manner in Section~\ref{sec: em}, and how to extract qualified new data in Section~\ref{sec: test-time}, respectively.

\subsection{Overview}
\label{sec: overview}
\begin{figure}[h]
  \centering
  \includegraphics[width=0.4\textwidth]{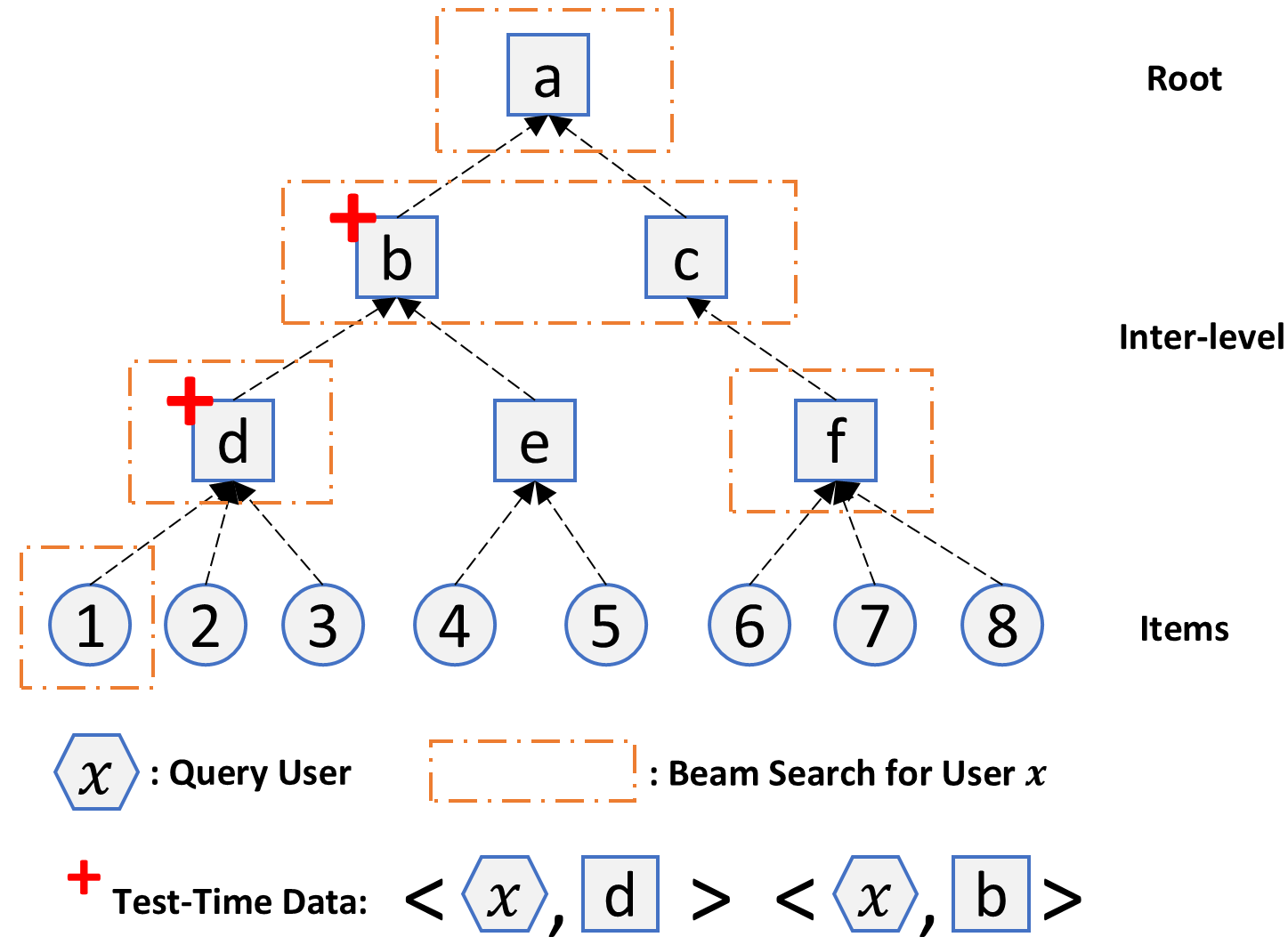}
  \caption{A Hierarchical Index Example Learnt by HILL.}
  \label{fig:HILL}
\end{figure}

In Figure~\ref{fig:HILL}, we show a hierarchical index example learnt by our HILL method.
To be more specific, in the illustration of Figure~\ref{fig:HILL}, we have 8 items ranging from 1 to 8.
The learnt hierarchical is a three layer tree, with node $a$ as the root. Given a user query $x$, starting from root $a$, a search algorithm (e.g., beam search with width as 2) locates item $1$ as the most relevant item for user $x$, which is included in the recommendation set for user $x$ and avoids the heavy computing of the similarity from each item to the user $x$.

More importantly, in Figure~\ref{fig:HILL}, tracing from the leaf node $1$ to the root node $a$, HILL also identifies a valuable but hidden path, i.e., $1$ $\rightarrow$ $d$ $\rightarrow$ $b$ $\rightarrow$ $a$, which records the interested inter-level nodes that user $x$ may also interest in. Since this tree is learnt, nodes $d$, $b$, $a$ are virtual nodes and not appear in the training data, which has the potential to fine-tune the model with user item pairs <$x, d$> and <$x, b$>. The reason HILL selected $d$ and $b$ but excluded $a$ is extended in Section 4.6.

\subsection{One-Layer Attention Learning}
\label{sec: one-layer}

To build up a hierarchical tree, the first fundamental step is to establish one layer, i.e., taking items as leaf nodes and mapping them to the upper level.   

Taking MoNN as an example to provide user and item embeddings, we next introduce our one-layer attention learning algorithm, as shown in Algorithm~\ref{alg:mapping_alg}, which takes the embedding vectors from a MoNN model as input and learns coarse index node embeddings for the item by minimizing the L2 distance between them. 
To be specific, our proposed algorithm employs an attention-based method by taking the item embedding as query, learnable embeddings for index nodes as keys and values, to calculate the index embeddings. 
With attention score, our algorithm allows the soft mapping during the training process, that is, one item can belong to multiple index nodes with varying weights. 

\begin{algorithm}[htbp!]
\caption{One-Layer Attention Learning}
\begin{algorithmic}[1]
\REQUIRE MoNN Model, Hyperparameter $K$
\ENSURE Item-Index Mapping Function $\bm{M}$, Embedding Matrix $\bm{R}$ of Index Nodes 
\STATE Randomly initialize index nodes' embedding vectors $\{\bm{c}_{k}\}_{k=1}^K$
\WHILE{not converge}
\Statex /* Mini-Batch Training */
    \FOR{each item $j$ in a batch}
        \STATE Sample a user-item pair <$i, j$> with label $y$, compute user embedding  $\bm{u}_i$ and item embedding $\bm{v}_j$ and through MoNN
        \STATE Compute the distance between item $j$ and index $k$ as \\ $d(j,k) = \|\bm{v}_j - \bm{c}_{k}\|^2$
        \STATE Compute the affinity between item $j$ and index $k$ as \\ $a_k =  \frac{e^{-\alpha * d(j,k)}}{\sum_{k'} e^{-\alpha * d(j,k')}}$
        \STATE Compute the pseudo item embedding by $\bar{\bm{c}} = \sum_k a_k \bm{c}_k$
        \STATE Update MoNN optimization with new pair $(y, \langle u_i, \bar{c} \rangle)$
    \ENDFOR
\Statex /* Finalize Index Embedding and Mapping*/
    \FOR{each $k$ index node}
    \FOR{each item $j$ in the corpora}
        \STATE Update $\bm{M}(j,k)$ by $d(j,k)$
        \STATE $\bm{c}_k$ = $\bm{v}_j$, if $\operatorname*{argmin}_j d(j,k)$
    \ENDFOR
    \ENDFOR
\ENDWHILE
\STATE Return function $\bm{M}$ and matrix $\bm{R}(k,:)$ = $\bm{c}_k$
\end{algorithmic}
\label{alg:mapping_alg}
\end{algorithm}

\subsection{Cross-Layer Residual Learning}
\label{sec: cross-layer}
Given that a single layer can be established, the next step is to stack it up iteratively. To make the index hierarchical, (i.e., the semantics of the upper level should depend on the lower level, and also store the information that lower level can not hold), inspired by residual quantization~\cite{lee2022autoregressive, zeghidour2021soundstream}, we aim to pass residue between input item embedding and the corresponding index node embedding at the lower level to its next index layer. 

Recall Lines 8--9 in Algorithm~\ref{alg:mapping_alg}, the index node representation learning is stored in the form (i.e., soft assignment or attention-based aggregation) of pseudo item embedding. It gives us the possibility of aligned dimensions across layers in the hierarchical tree. In other words, for each layer, we can have a ground truth item embedding and a pseudo item embedding that is closely associated with the index embedding.

Mathematically, given $K$ index nodes at each level\footnote{To simplify the notation, we denote each layer has the same number of index nodes, which is a hyperparameter during the implementation}, and the $N$ level to be established, our cross-layer residual learning can be expressed as follows:
Suppose we have the initial embedding vectors for all items by asking the MoNN model, and the embedding vector for item $j$ is denoted as $\bm{v}_{j}$;
Then, we can denote the initial residual vector $\bm{r}_{j}^{1}$ for item $j$ at the $1$-st layer as
\begin{equation}
    \bm{r}_{j}^{1} = \bm{v}_{j}
\end{equation}

According to Lines 8--9 in Algorithm~\ref{alg:mapping_alg}, HILL produces the pseudo item embedding vector $\bar{\bm{c}}_{j}^{1}$ for item $j$ when mapping the $1$-st layer to the $2$-nd layer. Hence, the gap for the residual learning to mix can be expressed as follows. At each level $n \in \{2,\ldots, N\}$, the recursive quantization of the residual vector is
\begin{equation}
    \bm{r}_{j}^{n} = \bm{r}_{j}^{n-1} - \bar{\bm{c}}_{j}^{n-1}
\end{equation}
The above residual vector $\bm{r}_{j}^{n}$ will serve as the item representation learning when building the next layer, i.e., replacing Line 4 in Algorithm~\ref{alg:mapping_alg} that asks MoNN to provide item embeddings.

Consequently, at each level $n$, the quantized embedding vector is computed as $\bm{q}_j^{n} = \sum_{l=1}^{n-1}{\bar{\bm{c}}_{j}^{l}}$.

The reconstruction loss $\mathcal{L}_{recon}$ is computed as:
\begin{align}
\mathcal{L}_{recon} = \sum_{j} \| \bm{q}_{j}^{N}-\bm{v}\|^2
\end{align}

The magnitude of residuals decreases when further moving down the hierarchy. As a result, a coarse index layer identifier expresses more general concepts, while a fine-grained index layer captures more detailed notions.

\subsection{Optimization}
\label{sec: optimization}
In order to better optimize our HILL algorithm for the index building, a few additional techniques are introduced to improve the training stability, including \textit{softmax temperature scheduler}, \textit{balanced index distribution}, and \textit{warmup strategy}.

\textbf{Softmax Temperature Scheduler}.
In the serving stage, the embedding vector of index nodes will serve as virtual items to help the user query to retrieve a bunch of its relevant items.
However, during the training process, our HILL algorithm uses a soft <item, index node> assignment, where each index node can be viewed as a combination of different interest items that can contain irrelevant items.
To mitigate the discrepancy, a scheduler is applied by gradually increasing the temperature (alpha), transitioning from soft assignment in the initial phase to a hard assignment later on. Small values of alpha yield a balanced distribution of the item-to-index assignment, while large values of alpha result in a skewed distribution. The scheduler is based on the following function:
\begin{align}
alpha = max\_alpha * \frac{current\_iter^{exp}}{max\_iters^{exp}}    
\end{align}

\textbf{Balanced Index Distribution}.
The index learning often suffers from cluster collapse, where the model utilizes only a limited subset of index nodes. A balanced index distribution is crucial to enable the use of neural network models with high complexity. Therefore, we employ the FLOPs regularizer to address this problem. The motivation stems from~\cite{paria2020minimizing}, which penalizes the model if all items are assigned to the same index node or if the distribution of <item, index node> assignment is imbalanced. Given it can be sensitive to smaller batches, data from the most recent $K$ batches is pooled and the FLOPs regularizer is applied on the pooled soft assignment matrix ($K$ * batch\_size, num\_index\_nodes).

\textbf{Warmup Strategy}.
A linear warm up strategy is employed for the index loss weight to gradually increase the learning rate, which is expected to stabilize the model parameters and mitigate the issue of item assignment oscillating between index nodes during the initial training phase.

\subsection{Expectation–Maximization Approximation by FAISS}
\label{sec: em}

The essence of learning the hierarchical index for a foundation model is to make the memory of the foundation model structurally organized, such that the relevance retrieval can use the optimal path and prune unnecessary branches. In the above sections, we already discussed how to learn the index via the co-training process. However, the co-training process inevitably involves the training of a foundation model, which can be infeasible during a short time window (e.g., hourly) and heavy workload (e.g., billion-scale users and million-scale items). In this viewpoint, we aim to propose an alternative learning method in case the computing resources are not adequate.

If we model the hierarchical index as an internal part of the well-trained foundation model, then we can solve the problem of learning the hierarchical index by using the Expectation-Maximization (EM) algorithm. To be more specific, if we view the index node as a (soft or hard) cluster of items, then we can model the item-index mapping as the hidden variables and user and item embedding as observation variables, such that the problem can be approximated as a Gaussian mixture model. Then, the E-step can be assigning each item its closest cluster and can be implemented by the GPU-enabled FAISS~\cite{DBLP:journals/tbd/JohnsonDJ21, DBLP:journals/corr/abs-2401-08281} parallel clustering library, e.g., K-Means; and the M-step can be training the foundation model according to current index embeddings. This iterative EM-based approximation enables fast optimization when computing resources are not necessary or for small updates between two time windows. In the experiments, we also show that this approximation leads to acceptable effectiveness.

Again, when computing resources are sufficient, a full version of HILL is recommended. The full version of HILL has advantages, including: (1) continuous training indexing with neural networks, which would be able to dynamically adapt to the latest data in the online streaming scenario; (2) avoiding the need to maintain different computational frameworks. The above conditions are important considerations for an industry-scale production system, to the best of our knowledge.

\subsection{New Data from HILL to Enable Test-Time Training}
\label{sec: test-time}

In the last part, we aim to introduce how, after establishing the hierarchical index, we discover the new qualified training data to fine-tune the foundation model.

As shown in Figure~\ref{fig:HILL}, the beam search from top to the bottom finds the item $1$ for user $x$.
If we traverse back, it is easy to identity a path: $1 \rightarrow d \rightarrow b \rightarrow a$, which implies that user $x$ is also interested in index nodes $d$, $b$, and $a$. 
Then, a natural question arises: can they make up new data pairs and help fine-tune the foundation model?
The answer is positive, since we can intuitively interpret the intermediate-level index nodes as the machine-readable category nodes for items after the (pre-)training process, and pairing them with original users generates the new training data <user, index node> (e.g., <user $u$, index $d$> and <user $u$, index $b$>), which is not seen in the previous training iterations and expects to bring new information to fine-tune the model. 

However, along this path, not all qualified intermediate-level index nodes are qualified enough to be selected to fine-tune the model. The first obvious clue is that the higher-level index nodes should be excluded from the new data set constructions, because the higher level an index node stands, the more general meaning it has. An extreme case is that the root node is shared by all users when we traverse back, such that adding <user, root> pair brings noise to the fine-tuning process. Therefore, the first hyperparameter for extracting the new data pairs, is how deep we traverse back from the bottom (i.e., item level), denoted as $\phi_{DEP}$. For example, in the above case, $\phi_{DEP}$ is 2 for getting the new data <user $u$, index $d$> and <user $u$, index $b$> for user $u$.

\begin{table*}[htbp!]
\centering
\caption{HILL-Enabled MoNN Performance with Baselines in Internal Data. I$_1$, I$_2$, V are in the order of O(1,000), O(100,000), O(10,000,000), M$_\text{S}$ denotes the model size of \underline{S}mall Model, and  M$_\text{L}$ denotes the model size of \underline{L}arge Model}
\begin{tabular}{l|c|c|c|c}
\toprule
\textbf{Model Architecture} & \textbf{Eval NE ($\downarrow$)} & \textbf{Recall ($\uparrow$)} & \textbf{Infra Cost ($\downarrow$)} & \textbf{Theoretical Cost} \\
\midrule
TTSN~\cite{DBLP:conf/nips/BromleyGLSS93} & baseline & 0\%     & 1x    & M$_{\text{XS}}$ × V \\
EBR~\cite{DBLP:conf/kdd/HuangSSXZPPOY20}  & +0.03\%  & -0.1\%  & 0.5x  & M$_{\text{XS}}$ × I$_1$ \\
MoNN Small        & -0.29\% & +2.4\%  & 2.5x  & M$_\text{S}$ × V \\
MoNN Medium       & -0.70\% & +4.2\%  & 17.3x & M$_\text{M}$ × V \\
MoNN Large        & -1.70\% & +9.4\%  & 24.6x & M$_\text{L}$ × V \\
\addlinespace
\makecell[l]{2-layer MoNN \\ (L$_1$: MoNN Small, L$_2$: TTSN)} & -0.23\% & +2.2\% & 1.7x & M$_\text{S}$ × I$_1$ + M$_{\text{XS}}$ × V \\
\addlinespace
\makecell[l]{2-layer MoNN \\ (L$_1$: MoNN Medium, L$_2$: MoNN Small)} & -0.47\% & +3.6\% & 3.3x & M$_\text{M}$ × I$_1$ + M$_\text{S}$ × V \\
\addlinespace
\makecell[l]{2-layer MoNN \\ (L$_1$: MoNN Large, L$_2$: MoNN Small)}  & -0.97\% & +6.0\% & 3.9x & M$_\text{L}$ × I$_1$ + M$_\text{S}$ × V \\
\bottomrule
\end{tabular}
\label{tab:monn-performance}
\end{table*}

Second, if we zoom in, index $d$ or index $b$ may also not qualify enough, because user $u$'s interest can spread over $d$'s items and other indices' items. We model that if user $u$'s interest concentrates only on $d$'s items, then <user $u$, index $d$> is a qualified new data pair. Mathematically, we propose an interest rate parameter $\phi_{IR}$, which takes a user $u$ and an index node $i$ at level $n$,
\begin{equation}
    \phi_{IR}(u,i^{n}) = \frac{|Int (u, n-1) \cap Child(i^{n})|}{Child(i^{n})} 
\end{equation}
where $Int (u, n-1)$ is a function to return the interested items (or index nodes) at level $n-1$ of the index tree, and $Child(i^{n})$ returns the set of direct children of node $i^{n}$ at level $n-1$.

Therefore, we can use $\phi_{IR}$ to filter out index nodes that only share a small portion of a certain user's interest, when we compose the fine-tuning data pairs for this user. An intuitive understanding of $\phi_{IR}$ can be: only if the user has very frequent preferences on <user, basketball> and <user, football>, and we discover <user, sports>; if user is interested in <user, basketball>, <user, music>, and <user, cooking>, then <user, sports> is not a strong signal. But again, in HILL, we did not dive deeper to learn to assign the human-readable semantic meaning for each index node, but only to use their embedding vectors.

Based on the above modeling, we discern that just a small $\phi_{DEP}$ and a large $\phi_{IR}(u,i^{n})$ threshold can select a small portion of new data, with which the fine-tuning can achieve a significant performance gain.
So far, we have only considered the positive strong signal in the above modeling, following the original loss during the fine-tuning. Moreover, weighing the new data pair with the weak signal and even negative signal, and adding contrastive learning loss functions during the fine-tuning process leave promising future directions.

\section{Experiments}
\label{sec: experiment}

In this section, we introduce the datasets, baselines, metrics, offline performance, and online service report.

\subsection{Datasets}
Here, we choose both public benchmark datasets and internal datasets to demonstrate the performance. For the public datasets, we select Gowalla, Yelp 2018, and Amazon-Book, as shown in Table~\ref{tab:dataset-statistics}, which records various user-item interactions and are publicly available\footnote{\url{https://github.com/PeiJieSun/NESCL}, \url{https://github.com/kuandeng/LightGCN}}.
The internal datasets are from daily Ads Recommendation tasks from Meta Platform.

\begin{table}[h!]
\centering
\caption{Statistics of Public Benchmark Datasets.}
\resizebox{0.465\textwidth}{!}{
\begin{tabular}{lrrrr}
\toprule
\textbf{Dataset} & \textbf{\# Users} & \textbf{\# Items} & \textbf{\# Interactions} & \textbf{Density} \\
\midrule
Gowalla     & 29,858  & 40,981 & 1,027,370 & 0.084\% \\
Yelp2018    & 31,688  & 38,048 & 1,561,406 & 0.130\% \\
Amazon-Book & 55,188  & 9,912  & 1,445,622 & 0.062\% \\
\bottomrule
\end{tabular}}
\label{tab:dataset-statistics}
\end{table}

\subsection{Baselines}
We select different categorical baselines, including (1) classic collaborative filtering methods, (2) neural collaborative filtering methods, (3) generative collaborative filtering methods, (4) industrial retrieval models, (5) general graph neural networks, and (6) graph neural network-inspired retrieval models. For the page limitation, we leave the reference of each baseline in Table~\ref{tab: public performance}.

\subsection{Metrics}
To verify the retrieval performance, metrics in the experiments consist of Recall, Normalized Discounted Cumulative Gain (NDCG), and Normalized Entropy (NE)

\textbf{Recall@$K$} measures the ability of a model to retrieve relevant items within the top-$K$ recommended list.
\begin{equation*}
\text{Recall@}K = \frac{|\text{Rel}_u \cap \text{Rec}_u^K|}{|\text{Rel}_u|}
\end{equation*}
where $\text{Rel}_u$ denotes the set of relevant (ground-truth) items for user $u$, and $\text{Rec}_u^K$ is the set of top-$K$ recommended items. Higher values indicate better coverage of relevant items. For the public datasets, we choose Recall@$20$.


\textbf{NDCG@K} accounts not only for the presence of relevant items in the recommendation list but also for their positions.
\begin{equation*}
\text{NDCG@}K = \frac{1}{|\mathcal{U}|} \sum_{u \in \mathcal{U}} \frac{\text{DCG}_u@K}{\text{IDCG}_u@K}
\end{equation*}
where $\text{DCG}_u@K = \sum_{i=1}^{K} \frac{\mathbb{I}(r_{u,i} = 1)}{\log_2(i+1)}$, $\mathbb{I}(y_{u,i} = 1)$ indicates whether the $i$-th item in the recommended list for user $u$ is relevant, and $\text{IDCG}_u@K = \sum_{i=1}^{\min(K, |\text{Rel}_u|)} \frac{1}{\log_2(i+1)}$. In other words, NDCG@K emphasizes recommending relevant items at higher ranks, and the higher the better.

\begin{table*}[htbp!]
\centering
\caption{Comparison of Baselines across Gowalla, Yelp2018, and Amazon-Book.}
\scalebox{1}{%
\begin{tabular}{l|cc|cc|cc}
\toprule
\textbf{Baseline} & \multicolumn{2}{c|}{\textbf{Gowalla}} & \multicolumn{2}{c|}{\textbf{Yelp 2018}} & \multicolumn{2}{c}{\textbf{Amazon-Book}} \\
 & Recall@20 & NDCG@20 & Recall@20 & NDCG@20 & Recall@20 & NDCG@20 \\
\midrule
BPR~\cite{DBLP:conf/uai/RendleFGS09} & 0.1627 & 0.1378 & 0.0576 & 0.0468 & 0.0338 & 0.0261 \\
GRMF~\cite{DBLP:conf/nips/RaoYRD15} & 0.1477 & 0.1205 & 0.0571 & 0.0462 & 0.0354 & 0.0270  \\
GRMF-norm~\cite{DBLP:conf/nips/RaoYRD15} & 0.1557 & 0.1261 & 0.0561 & 0.0454 & 0.0352 & 0.0269  \\
HOP-Rec~\cite{DBLP:conf/recsys/YangCWT18} & 0.1399 & 0.1214 & 0.0517 & 0.0428 & 0.0309 & 0.0232  \\
ENMF~\cite{DBLP:journals/tois/ChenZZLM20} & 0.1523 & 0.1315 & 0.0624 & 0.0515 & 0.0359 & 0.0281 \\
MF-CCL~\cite{DBLP:conf/cikm/MaoZWDDXH21} & 0.1837 & 0.1493 & 0.0698 & 0.0572 & 0.0559 & 0.0447  \\
SimpleX~\cite{DBLP:conf/cikm/MaoZWDDXH21} & 0.1872 & 0.1557 & 0.0701 & 0.0575 & 0.0583 & 0.0468 \\
NeuMF~\cite{DBLP:conf/sigir/Wang0WFC19} & 0.1399 & 0.1212 & 0.0451 & 0.0363 & 0.0258 & 0.0200 \\
Mult-VAE~\cite{DBLP:conf/www/LiangKHJ18} & 0.1641 & 0.1335 & 0.0584 & 0.0450 & 0.0407 & 0.0315 \\
Macrid-VAE~\cite{DBLP:conf/nips/MaZ0Y019} & 0.1618 & 0.1202 & 0.0612 & 0.0495 & 0.0383 & 0.0295  \\
YouTubeNet~\cite{DBLP:conf/recsys/CovingtonAS16} & 0.1754 & 0.1473 & 0.0686 & 0.0567 & 0.0502 & 0.0388  \\
CMN~\cite{DBLP:conf/sigir/EbesuSF18} & 0.1405 & 0.1221 & 0.0475 & 0.0369 & 0.0267 & 0.0218 \\
CML~\cite{DBLP:conf/www/HsiehYCLBE17} & 0.1670 & 0.1292 & 0.0622 & 0.0536 & 0.0522 & 0.0428  \\
DeepWalk~\cite{DBLP:conf/kdd/PerozziAS14} & 0.1034 & 0.0740 & 0.0476 & 0.0378 & 0.0346 & 0.0264  \\
LINE~\cite{DBLP:conf/www/TangQWZYM15} & 0.1335 & 0.1056  & 0.0549 & 0.0446 & 0.0410 & 0.0318  \\
Node2Vec~\cite{DBLP:conf/kdd/GroverL16} & 0.1019 & 0.0709 & 0.0452 & 0.0350 & 0.0402 & 0.0309 \\
Item2Vec~\cite{DBLP:conf/mlsp/BarkanK16} & 0.1325 & 0.1057 & 0.0503 & 0.0411 & 0.0326 & 0.0251\\
GAT~\cite{DBLP:conf/iclr/VelickovicCCRLB18} & 0.1401 & 0.1401 & 0.0543 & 0.0431 & 0.0326 & 0.0235  \\
JKNet~\cite{DBLP:conf/icml/XuLTSKJ18} & 0.1622 & 0.1391 & 0.0608 & 0.0502 & 0.0268 & 0.0343  \\
DropEdge~\cite{DBLP:conf/iclr/RongHXH20} & 0.1627 & 0.1394 & 0.0614 & 0.0506 & 0.0342 & 0.0270  \\
APPNP~\cite{DBLP:conf/iclr/KlicperaBG19} & 0.1708 & 0.1462  & 0.0635 & 0.0521 & 0.0384 & 0.0299  \\
DisenGCN~\cite{DBLP:conf/icml/Ma0KW019} & 0.1356 & 0.1174 & 0.0558 & 0.0454 & 0.0329 & 0.0254  \\
LightGCN~\cite{DBLP:conf/sigir/0001DWLZ020} & 0.1830 & 0.1554 & 0.0649 & 0.0530 & 0.0411 & 0.0315 \\
GC-MC~\cite{DBLP:journals/corr/BergKW17} & 0.1395 & 0.1204 & 0.0462 & 0.0379 & 0.0288 & 0.0224  \\
PinSage~\cite{DBLP:conf/kdd/YingHCEHL18} & 0.1380 & 0.1196 & 0.0471 & 0.0393 & 0.0282 & 0.0219 \\
NIA-GCN~\cite{DBLP:conf/sigir/SunZGGTHMC20} & 0.1359 & 0.1106 & 0.0599 & 0.0491 & 0.0369 & 0.0287  \\
SGL-ED~\cite{DBLP:conf/sigir/WuWF0CLX21} & 0.1835 & 0.1539 & 0.0675 & 0.0555 & 0.0478 & 0.0379  \\
DeosGCF~\cite{DBLP:conf/tagml/LiuMJZY22} & 0.1784 & 0.1477 & 0.0626 & 0.0504 & 0.0410 & 0.0316  \\
IMP-GCN~\cite{DBLP:conf/www/LiuCZGN21} & 0.1845 & 0.1567  & 0.0653 & 0.0531 & 0.0460 & 0.0357  \\
BUIR~\cite{DBLP:conf/sigir/LeeKJPY21} & 0.1575 & 0.1301  & 0.0647 & 0.0526 & 0.0439 & 0.0346  \\
DGCF~\cite{DBLP:conf/sigir/WangJZ0XC20} & 0.1842 & 0.1561 & 0.0654 & 0.0534 & 0.0422 & 0.0324  \\
IA-GCN~\cite{DBLP:journals/corr/abs-2204-03827} & 0.1839 & 0.1562  & 0.0659 & 0.0537 & 0.0472 & 0.0373 \\
LT-OCF~\cite{DBLP:conf/cikm/0002JP21} & 0.1875 & 0.1574 & 0.0671 & 0.0549 & 0.0442 & 0.0341  \\
HMLET~\cite{DBLP:conf/wsdm/KongKJ0LPK22} & 0.1874 & 0.1589 & 0.0675 & 0.0557 & 0.0482 & 0.0371  \\
GTN~\cite{DBLP:conf/sigir/FanL0ZT022} & 0.1870 & 0.1588 & 0.0679 & 0.0554 & 0.0450 & 0.0346  \\
MGDCF~\cite{DBLP:journals/tkde/HuHQFX24} & 0.1864 & 0.1589 & 0.0696 & 0.0572 & 0.0490 & 0.0378  \\
BSPM-LM~\cite{DBLP:conf/sigir/00020PC23} & 0.1901 & 0.1570 & 0.0713 & 0.0584 & \textbf{0.0733} & \textbf{0.0610}  \\
NESCL~\cite{DBLP:journals/tkde/SunWZCW24} & \underline{0.1908} & \underline{0.1614} & \underline{0.0740} & \underline{0.0609} & 0.0623 & 0.0509 \\
\textbf{HILL} (Ours) & \textbf{0.1924} & \textbf{0.1628} & \textbf{0.0745} & \textbf{0.0612} & \underline{0.0625} & \underline{0.0513} \\
\bottomrule
\end{tabular}%
}
\label{tab: public performance}
\end{table*}

\textbf{NE} is selected based on the previous recommendation lessons at Facebook~\cite{DBLP:conf/kdd/HePJXLXSAHBC14}. Assume a given training data set has $N$ examples with labels $y_i \in  \{-1, +1\}$ and estimated probability of click $p_i$ where $i = 1, 2, \ldots, N$. The average empirical
CTR as $p$, then
\begin{equation}
NE = \frac{-\frac{1}{N} \sum_{i=1}^{n} \left( \frac{1 + y_i}{2} \log(p_i) + \frac{1 - y_i}{2} \log(1 - p_i) \right)}{ - (p \cdot \log(p) + (1 - p) \cdot \log(1 - p)) }
\end{equation}
The reason for this normalization is that the closer the background CTR is to either 0 or 1, the easier it is to achieve a better log loss.
Dividing by the entropy of the background CTR makes the NE insensitive to the background CTR.
The lower the value, the better the prediction made by the model.

\begin{equation}
\text{Normalized Entropy@}K = \frac{-\sum_{i \in \mathcal{I}} p_i \log p_i}{\log |\mathcal{I}_K|}
\end{equation}

\subsection{Offline Performance}
Here, the performance will be demonstrated in two aspects, i.e., internal test, public benchmark, and ablation study to verify the important components.

\textbf{Internal Test}.
The effectiveness and efficiency analysis of MoNN with baselines are shown in Table~\ref{tab:monn-performance}, where all MoNN models are equipped with joint optimization of HILL and show significant performance, e.g., $>0.05\%$ NE Gain is significant~\cite{DBLP:conf/kdd/HePJXLXSAHBC14}.

To be specific, in Table~\ref{tab:monn-performance}, by scaling up the model size, we can observe that MoNN Large performs the best, in terms of NE and Recall metrics. However, it brings considerable infrastructure cost. Therefore, we propose to stack MoNN blocks and make the lower level take the majority (but not all) of the data during the training
. As shown in the last three rows of Table~\ref{tab:monn-performance}, which largely reduces the infrastructure cost and keeps the competitive effectiveness. In addition to theoretical efficiency analysis, we present the cost of serving the MoNN model based on the following parameters: \textbf{I$_1$} (number of nodes in L$_1$ layer), \textbf{I$_2$} (number of nodes in L$_2$ layer) and \textbf{V} (number of items in the corpus), \textbf{M$_{XS}$} denotes the cost to serve Two Tower model, \textbf{M$_S$} denotes the cost to serve MoNN Small, \textbf{M$_M$} the cost to serve MoNN Medium, \textbf{M$_L$} the cost to serve MoNN Large.

\textbf{Ablation Studies}.
After showing that HILL can serve the large retrieval model effectively and efficiently, we then execute ablation studies to verify our theoretical design.

First, we design three training tricks during HILL optimizations, i.e., Softmax Temperature Scheduler, Balanced Index Distribution, Warmup Strategy.
In Table~\ref{tab: training tricks}, we observe that (1) the full version of including all training tricks has the best performance; (2) removing each one can reduce the performance, and they do not conflict with each other; (3) the Softmax Temperature Scheduler has the most significant loss when it is removed.

\begin{table}[htbp!]
\caption{Ablation Study of Training HILL, MoNN Small as Backbone, on Gowalla Dataset.}
\begin{tabular}{c|c}
    \toprule
    \textbf{Training Variant} & \textbf{NE ($\downarrow$)} \\
    \midrule
    w/o Softmax Temperature Scheduler & +0.10\% \\
    w/o Balanced Index Distribution & +0.05\% \\
    w/o Warmup Strategy & +0.03\% \\
    \bottomrule
\end{tabular}
\label{tab: training tricks}
\end{table}

Second, in Table~\ref{tab: em}, we show that the EM version of HILL can also achieve a competitive performance, given the NE loss is less than $0.04\%$, which also suggests that the full version of HILL has a fair reason to be considered when the computing resource allows.

\begin{table}[htbp!]
\caption{Ablation Study of HILL Approximation, 2-Layer MoNN as Backbone, on Gowalla Dataset.}
\begin{tabular}{c|c}
    \toprule
    \textbf{Model Architecture} & \textbf{NE ($\downarrow$)} \\
    \midrule
    HILL & -0.15\% \\
    HILL (EM) & -0.11\% \\
    \bottomrule
\end{tabular}
\label{tab: em}
\end{table}

\textbf{Public Benchmark}.
In addition to the internal datasets, we also conduct the experiments on the public dataset to show the effectiveness of HILL by choosing NESCL~\cite{DBLP:journals/tkde/SunWZCW24} as the retrieval model and learning the corresponding hierarchical index. To be specific, we first use the EM version of HILL to learn the hierarchical index; then, we extract the <user, index node> data pairs in the index to fine-tune the retrieval model, and finally report the performance in Table~\ref{tab: public performance}. For example, in the Gowalla dataset, the outperformance is obtained by setting $\phi_{DEP} = 2$, $phi_{IR} = 0.8$ at the second layer, and $phi_{IR} = 0.4$ at the third layer.

\textbf{Parameter Analysis}.
Then, a natural question arises: whether including more new data pairs can further improve the performance? To answer this question, we prepare the parameter analysis in Tables~\ref{tab:parameter_analysis_1} and \ref{tab:parameter_analysis_2}, which shows that the small amount but precise test-time data is sufficient for the leading performance.

\begin{table}[ht]
\centering
\caption{Performance with different $\phi_{\text{DEP}}$.}
\scalebox{0.8}{
\begin{tabular}{ccccc}
\toprule
$\phi_{\text{DEP}}$ & \# Nodes per Inter-layer & $\phi_{\text{IR}}$ & Recall@20 & NDCG@20 \\
\midrule
1 & 8000                      & 0.8                      & 0.1922 & 0.1624 \\
2 & 8000, 800                 & 0.8, 0.4                 & \textbf{0.1924} & \textbf{0.1628} \\
3 & 8000, 800, 80             & 0.8, 0.4, 0.2            & 0.1922 & 0.1624 \\
4 & 8000, 8000, 80, 8         & 0.8, 0.4, 0.2, 0.1       & 0.1916 & 0.1623 \\
\bottomrule
\end{tabular}}
\label{tab:parameter_analysis_1}
\end{table}

\begin{table}[ht]
\centering
\caption{Performance with different $\phi_{\text{IR}}$.}
\scalebox{0.9}{
\begin{tabular}{ccccc}
\toprule
$\phi_{\text{DEP}}$ & \# Nodes per Inter-layer & $\phi_{\text{IR}}$ & Recall@20 & NDCG@20 \\
\midrule
1 & 8000, 800 & 0.8, 0.1 & 0.1916 & 0.1624 \\
1 & 8000, 800 & 0.8, 0.2 & 0.1923 & \textbf{0.1629} \\
1 & 8000, 800 & 0.8, 0.3 & 0.1914 & 0.1624 \\
1 & 8000, 800 & 0.8, 0.4 & \underline{0.1924} & \underline{0.1628} \\
1 & 8000, 800 & 0.8, 0.5 & 0.1914 & 0.1623 \\
1 & 8000, 800 & 0.8, 0.6 & 0.1924 & 0.1620 \\
1 & 8000, 800 & 0.8, 0.7 & \textbf{0.1925} & 0.1623 \\
\bottomrule
\end{tabular}}
\label{tab:parameter_analysis_2}
\end{table}

\subsection{Online Service Report}
\label{sec:online_service}
The MoNN foundation retrieval model has been successfully deployed at Meta for Ads Retrieval. Prior to the deployment of MoNN Large, we first launched the MoNN Small architecture to production, given the relatively low infra cost. As shown in Table~\ref{tab:online}, online A/B tests demonstrate that 2-Layer MoNN led to \textbf{2.57\%} online ads metric gains.

\begin{table}[h]
\caption{Online performance in Meta Ads Production.}
\begin{tabular}{c|c}
    \toprule
    \textbf{Model Architecture} & \textbf{\makecell{Online Metric ($\uparrow$)}} \\
    \midrule
    TTSN & -0.21\% \\
    \hline
    MoNN Small & baseline \\
    \hline
    \makecell{2-Layer MoNN \\ (L$_1$: MoNN Medium, L$_2$: MoNN Small)} & +1.22\% \\
    \hline
    \makecell{2-Layer MoNN \\ (L$_1$: MoNN Large, L$_2$: MoNN Small)} & +2.57\% \\
    \bottomrule
\end{tabular}
\label{tab:online}
\end{table}

\subsection{Details of Deploying MoNN}
\label{sec: MoNN details}
\begin{figure}[htbp]
  \centering
  \includegraphics[width=0.4\textwidth]{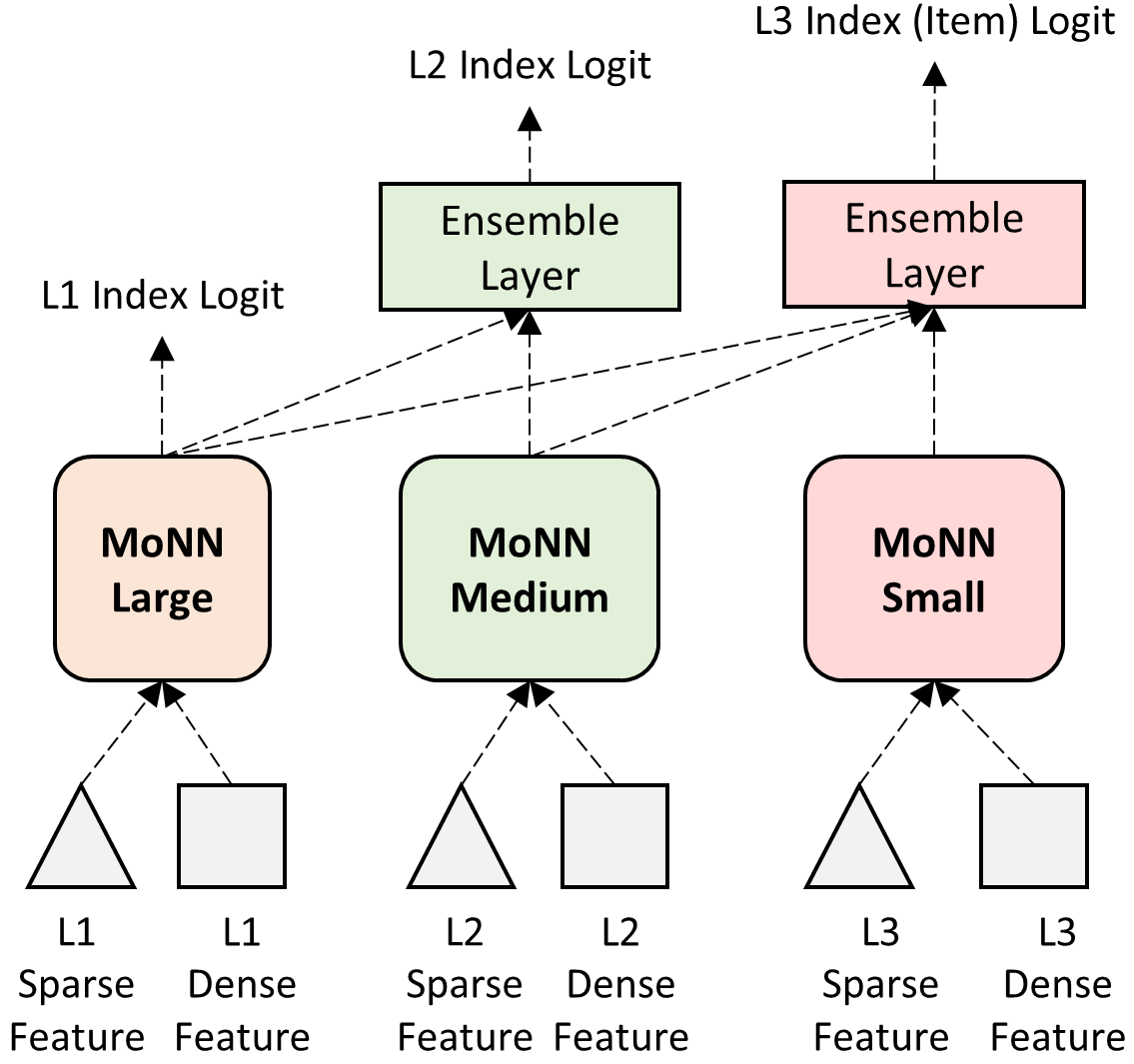}
  \caption{3-Layer MoNN Illustration Example.}
  \label{fig:hsnn}
\end{figure}

In this section, we introduce the stacked MoNN model.
Figure~\ref{fig:hsnn} shows a 3-layer MoNN model architecture: L$_1$ layer, L$_2$ layer, and L$_3$ layer.
The L$_1$ layer operates on the coarsest index granularity and it is able to take advantage of a MoNN model architecture with the highest complexity (through computation sharing) and a wide range of interaction features (<user, L$_1$ index node>).
Then, the L$_3$ index operates on the finest granularity (in extreme case, it can be item level directly) and leverages a MoNN block with the lowest complexity.
The three MoNN modules are combined using the ensemble layer. This manner allows final prediction to leverage multiple MoNNs with different model complexity to consume different granularity of features, resulting in a more accurate prediction.

\textbf{Features}. MoNN Small processes features at the individual item level, utilizing user features, item features, and <user, item> interaction features. In contrast, MoNN Medium and MoNN Large operate at a coarser granularity (L$_2$ layer and L$_1$ layer, respectively) and consume user features, index node features, and <user, index node> interaction features, where index node is replaced by a representative item for feature computation.

\textbf{Loss Function}. Each layer of multi-layer MoNN has its own loss function, i.e., with 1 <user, item> prediction loss function and <user, index node> prediction loss function.

\section{Related Work}
\label{sec: related work}
To support retrieval models, prior works suggest that organizing candidate items into index structures can reduce the search space and expedite the identification of relevant item-user pairs~\cite{DBLP:conf/kdd/ZhuLZLHLG18, gao2020deep, DBLP:conf/nips/FengLLLC22, DBLP:conf/sigir/LiAZM0LC23, liu2024learning, DBLP:conf/iclr/Li0LLBY00G25}. However, such methods are often inadequate for modern industrial-scale foundation models, which typically employ deep, densely connected neural architectures to capture complex user-item interactions enriched with structured contextual information. To the best of our knowledge, existing indexing techniques fall short of addressing the exactness and scalability demands of these models.
Motivated by this limitation, we propose to learn a hierarchical index tailored to the memory components of large-scale foundation retrieval models. This index is designed to support exactness-aware search, enabling efficient inference by bypassing redundant search paths and thereby reducing unnecessary computational overhead.

Meanwhile, the notion of scaling laws at training has recently been extended to the context of foundation models for ranking and retrieval~\cite{DBLP:journals/corr/abs-2001-08361, DBLP:journals/corr/abs-2203-15556, DBLP:journals/corr/abs-2208-08489, DBLP:conf/aaai/ShinKKRJ0K23, DBLP:conf/recsys/ZhangHLCZW24, DBLP:conf/sigir/FangZAMS0024, DBLP:journals/corr/abs-2412-00714}. Test-time training strategies in recommendation systems remain largely underexplored.
Our work aims to address this gap and spark further research into efficient and scalable inference for foundation retrieval models.

\section{Conclusion}
\label{sec: conclusion}
To make the large-scale foundation retrieval model serve effectively and efficiently, in this paper, we propose the hierarchical index learning method HILL to learn the index structure over the memory of the foundation model, taking MoNN (i.e., a deployed retrieval model at Meta for Ads Retrieval) for illustration. Moreover, we found that learnt index convey a small set of new and high-quality data pairs that can be used to test-time fine-tune the model to boost the performance.

\bibliographystyle{ACM-Reference-Format}
\bibliography{reference}

\end{document}